\documentclass[aps,english,prb,twocolumn,showpacs,superscriptaddress]{revtex4}
\usepackage{graphicx}

\begin{document}

\title{High-performance planar nanoscale dielectric capacitors}

\author{V. Ongun \"{O}z\c{c}elik}
\affiliation{UNAM-National Nanotechnology Research Center, Bilkent University, 06800 Ankara, Turkey}
\affiliation{Institute of Materials Science and Nanotechnology, Bilkent University, Ankara 06800, Turkey}
\author{S. Ciraci}
\affiliation{Department of Physics, Bilkent University, Ankara 06800, Turkey}

\begin{abstract}

We propose a model for planar nanoscale dielectric capacitor consisting of a single layer, insulating hexagonal boron nitride (BN) stripe placed between two metallic graphene stripes, all forming commensurately a single atomic plane. First-principles density functional calculations on these nanoscale capacitors for different levels of charging and different widths of graphene - BN stripes mark high gravimetric capacitance values, which are comparable to those of supercapacitors made from other carbon based materials. Present nanocapacitor model allows the fabrication of series, parallel and mixed combinations which offer potential applications in 2D flexible nanoelectronics, energy storage and heat-pressure sensing systems.

\end{abstract}

\maketitle

\section{Introduction}
While sustainable clean energy production and storage have been one of the critical concerns of our century, nanoscale capacitors are developing as an energy storage medium due to their recyclable high-energy storage based on charge-separation.\cite{stoller2008graphene} In particular, their flexibility, light weight and high performance have made nanoscale capacitors advantageous in the field of electronics and energy storage. With this regard, graphene with its one-atom thick layer and  high chemical stability \cite{novoselov2005two, geim2007rise} has been recently used in electrochemical capacitors. \cite{stoller2008graphene,si2008exfoliated, vivekchand2008graphene,yoo2011ultrathin,wang2009supercapacitor,pandolfo2006carbon} As opposed to electrochemical double-layer capacitors or supercapacitors, where an aqueous solution is used to store electric energy with electron transfer through chemical reactions, a nanoscale dielectric capacitor (NDC) uses a solid dielectric between the metallic plates and can attain properties that are superior to other systems used in energy storage. NDCs are favored in various applications, such as hybrid devices and personal electronics due to their short load cycle and longer life. These properties have even greater advantage where flexibility and smaller sizes are desired.\cite{simon2008materials,chen2009flexible,yu2010ultrathin}

\begin{figure}
\includegraphics[width=8cm]{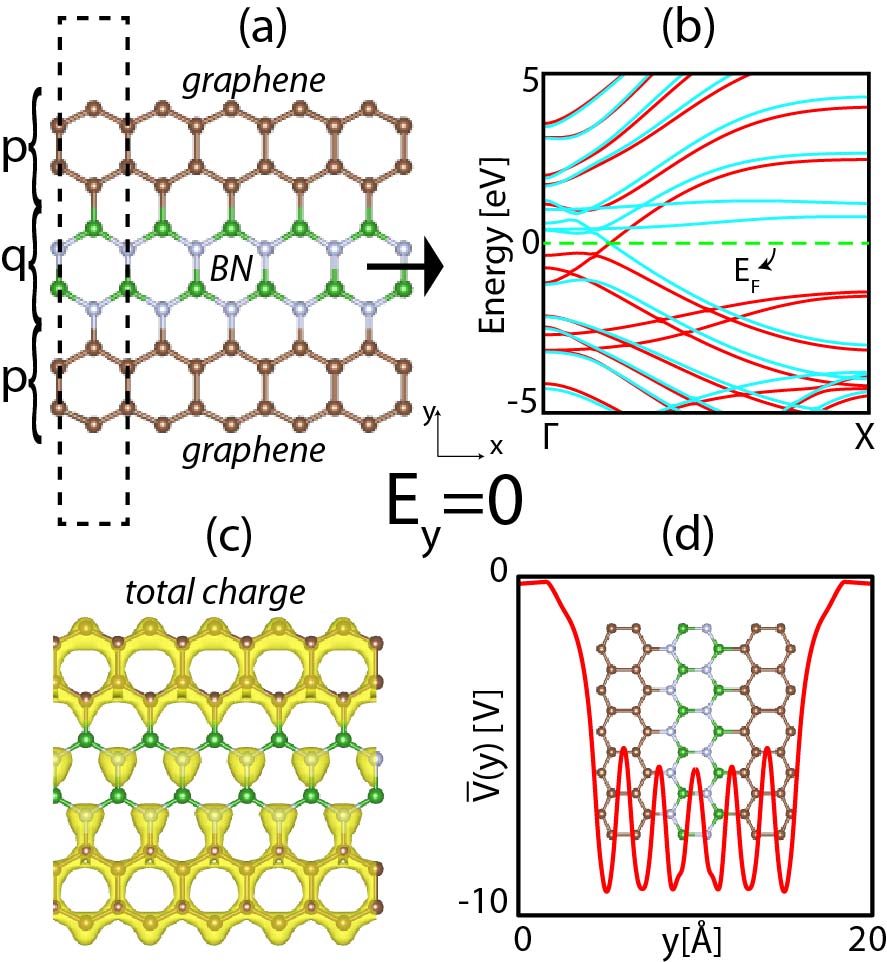}
\caption{(a) PNDC formed by a zigzag BN stripe placed between metallic graphene stripes, which display 1D translational periodicity along $x$-direction. The unit cell is delineated by dashed lines. $p$ and $q$ are number of atoms in graphene and BN stripes in the unit cell. (b) Electronic band structure of PNDC under zero bias voltage (or $E_y$=0), where spin up and down states are shown with red(dark) and blue(light) lines, respectively. (c) Total charge density $\rho(\textbf{r})$ isosurfaces of PNDC. (d) $(xz)$- plane averaged electronic potential, $\bar{V}(y)$.}
\label{fig1}
\end{figure}

\begin{figure*}
\includegraphics[width=14cm]{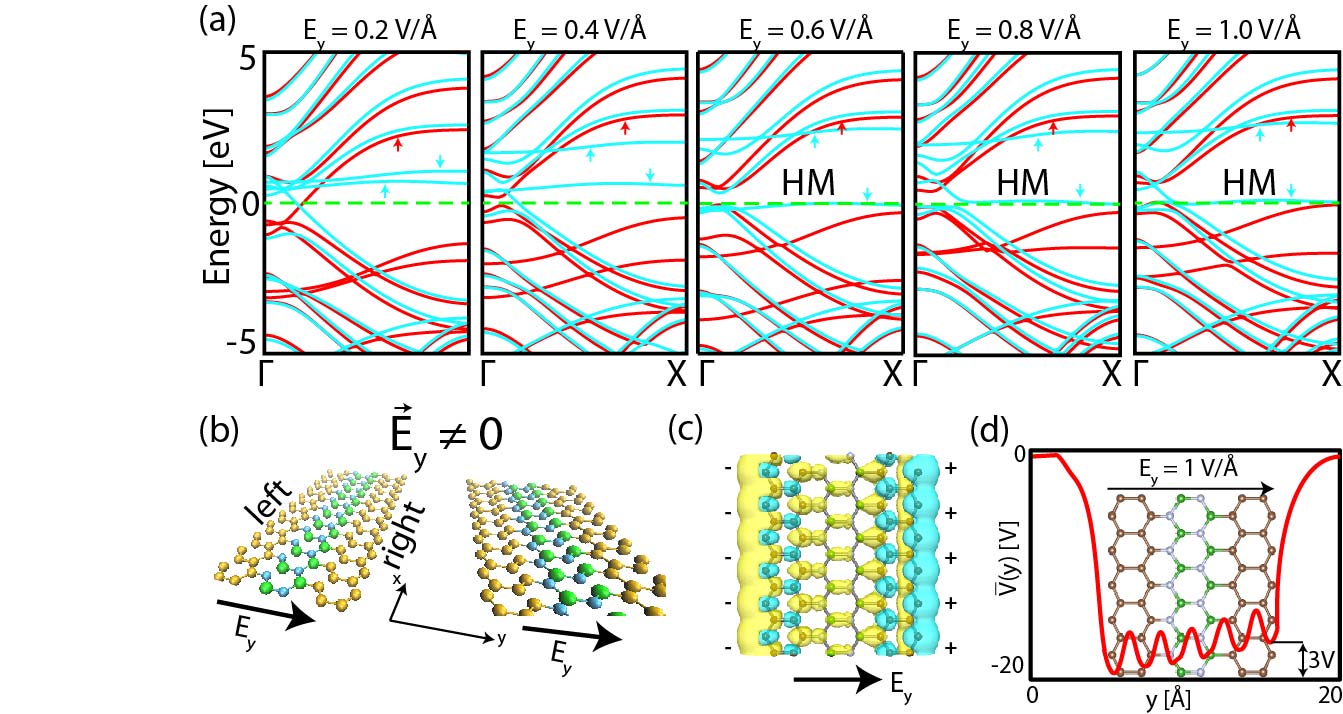}
\caption{(a) Evolution of the energy bands of PNDC[4/4/4] under applied electric field $\vec{E_y}$. Shifts of relevant spin bands are indicated by arrows. For $E_{y} \geq$ 0.6 V/\AA, PNDC becomes a half-metal (HM). (b) A perspective view of the array of PNDC[4/4/4] periodically repeating along $y$-axis. (c) The isosurface of difference charge density, $\Delta (\rho)$, showing the charge separation, where the right graphene stripe is depleted from electrons, which are in turn deposited to left graphene stripe due to the shifts of bands under $\vec{E_y}$. (d) $(xz)$-plane averaged electronic potential $\bar{V}(y,\vec{E_y})$ exhibiting a potential difference of $\Delta \bar{V}(y)$=3V under $E_y$=1 V/ \AA.}

\label{fig2}
\end{figure*}

Since singe layer, hexagonal boron nitride (BN) can be grown on top of graphene, \cite{ozccelik2012epitaxial, sachs2011adhesion, liu2011direct} the use of BN as the solid dielectric nanoscale precision spacer between graphene layers is promising candidate for the fabrication of graphene based NDCs. On the other hand, graphene layers having perfect electron-hole symmetry can function as metallic plates to store and release charge.  Earlier, based on first-principles calculations, we predicted that an NDC consisting of a few insulating BN layers placed between two lattice-matched graphene layers can achieve high capacitance values. \cite{ozcelik2013nanoscale} Furthermore, in that NDC model the capacitive behavior of the system diverges from the classical Helmholtz model as the separation between the metallic plates gets smaller, due to the increase in the dielectric constant of BN thin films. This anomalous size-dependent increase of the capacitance was also experimentally verified soon after our prediction. \cite{shi2014boron} Also it has been experimentally demonstrated that the permittivity of BN thin films increase significantly and leads to an increase in the capacitance value much higher than what is attained classically.

In this study, we show that \textit{planar nanoscale dielectric capacitors} (PNDCs) can be realized as a one-atom-thick, single layer honeycomb structure consisting of a BN stripe as dielectric spacer between two metallic graphene stripes. These laterally stacked stripes are lattice matched. Furthermore, using these PNDCs, one can achieve high energy storage, as well as high gravimetric capacitance values, which are comparable to those of supercapacitors. As an alternative to NDCs consisting of vertically stacked dielectric layers capped by two metallic graphene, PNDCs achieve charge separation between two separated, parallel metallic stripes embedded in the same atomic plane.

It was shown experimentally \cite{ci2010atomic,lim2014stacking,drost2014electronic} that monolayer metallic graphene and dielectric BN can be grown commensurately with desired periodicity and with sharp boundaries. Thus one can construct not only single planar capacitors but also various series and parallel combinations of the former to achieve higher potential difference or charge separation. Thus PNDCs predicted in the present study will make an important forward step towards the fabrication of atomically thin circuitry based on graphene/BN lateral heterostructures \cite{rubio2012, levendorf2012graphene} to allow microwave, as well as heat and pressure sensing applications.

\section{Method}
Due to nanoscale sizes of the stripes, the stored energy through charge separation has to be treated from the first-principles. However, available first-principles methods allow us to treat only one kind of uniform excess charge (negative or positive) in the same system at a time. The crucial aspect of this study is to maintain the charge separation of opposite polarity in two graphene stripes at both sides of BN by applying an external electric field, $\vec{E_y}$, in the plane of PNDC but perpendicular to the axis of the stripe. However, the electronic potential under the applied electric field makes a dip in the vacuum spacing between PNDC stripes treated within the periodic boundary conditions using supercell geometry. As the strength of the electric field increases, this dip is further lowered as if a quantum well and allows the plane wave basis set to have states confined to the well as the solution of the Hamiltonian. Once the energies of these confined states are lowered below the Fermi level they start to be occupied by electrons. As a result, the electrons in the graphene layers are going to spill into the vacuum region under the external applied electric field. The spilling of the charge from PNDC to the vacuum is clearly erroneous and unrealistic. These artifacts of plane wave basis set become even more critical for wide vacuum spacing. On the other hand narrow vacuum spacing is not desired since it gives rise to significant coupling between adjacent PNDC treated using the supercell geometry. These artifacts can be eliminated by using basis set consisting of orbitals centered only at the atomic sites, which does not allow the states confined to the potential dip in the vacuum spacing.\cite{topsakal2013effects, gurel2013effects} Thus, we carried out first-principles, spin-polarized calculations within density functional theory, where the eigenstates of Kohn-Sham Hamiltonian are expressed as linear combination of numerical atomic orbitals. The exchange-correlation potential was approximated by Perdew, Burke and Ernzerhof  functional. \cite{pbe} A 200 Ryd mesh cut-off was chosen. Core states were replaced by norm-conserving, non-local Trouiller-Martins pseudopotentials. \cite{troullier1991} Atomic positions and lattice constants were optimized using the conjugate gradient method by minimizing the total energy and atomic forces for each configuration. Dipole corrections \cite{payne} were applied in order to remove spurious dipole interactions. All numerical calculations were performed using the SIESTA
code.\cite{siesta}

\section{Results and Discussions}
Our model of PNDC described in Fig. \ref{fig1}(a) is composed of lateral stacking of two narrow, zigzag graphene nanoribbons (stripes), ZGNR(p) and a zigzag BN nanoribbon, ZBNR(q) in between. Here p and q indicate the number of C and B-N atoms in their unit cells, respectively. Bare ZGNR(p) and ZBNR(q) are metallic. When saturated by hydrogen atoms, ZBNR(q) nanoribbon transforms into an insulator. \cite{topsakalBN} Similarly, when joined to graphene nanoribbons commensurately at both sides, the same ZBNR(q) can function as a dielectric spacer, while the graphene nanoribbons continue to be metal. These stripes can be set to desired widths by changing p and q values. While these three lattice matched graphene/BN/graphene composite stripes preserve their translational periodicity along $x$-axis and acquire 1D character, the translational symmetry is broken along $y$-direction. This model is treated by using periodic boundary conditions, where 1D PNDCs are repeated periodically along $y$- and $z$-axis keeping 10 \AA~ vacuum spacing between them. The electronic structure of PNDC[p/q/p] structure under zero bias $\Delta V$=0 (or $\vec{E_y}$=0) is shown in Fig. \ref{fig1}(b). It has spin polarized ground state with magnetic moment $\mu$= 2.3 $\mu_B$. Graphene-like and BN-like bands are separated in the direct space and Fermi level is pinned at graphene sides while BN remains as an insulator. As seen in Fig. \ref{fig1}(c), owing to the different ionicities of C, N, and B; a minute charge is transferred across the zigzag interface even for $\vec{E_y}$=0. However, this does not lead to a capacitive behavior. In fact, the ($xz$)-plane averaged electronic potential $\bar{V}(y)$ in Fig. \ref{fig1}(d) indicates a vanishing potential difference, i.e. $\Delta \bar{V}(y)$=0 between graphene stripes.

Normally, when connected to a real electric circuit, separation of charges of opposite polarity is attained between graphene stripes of PNDC under the applied external bias voltage $\Delta V$, whereby the chemical potentials at metallic graphene stripes shift by e$\Delta V$. In this non-equilibrium state an electric field is generated across the BN stripe. Here for the reason explained above, we simulate this normal operation of PNDC by applying an electric field $\vec{E_y}$ along the $y$-direction with magnitude in the range of 0 V/\AA~ to 1 V/\AA. Notably, even if $E_y$=1 V/\AA~ appears to be high, in earlier studies such high field values were attained. \cite{highfield} For each value of applied $\vec{E_y}$, we first carried out calculations to optimize the atomic structure and lattice constants of PNDC by minimizing the total energy and atomic forces. Also, for each value of applied $\vec{E_y}$ we calculated the electronic energy bands of optimized structures as well as the net charge on the left graphene stripe, $-Q$ in e/cell. This is obtained from the integral

\begin{equation}
-Q = \int \Delta \rho(\textbf{r}, \vec{E_y})d\textbf{r}
\label{integral}
\end{equation}
computed in the part of the PNDC unit cell corresponding to the left graphene stripe, where the boundary between the left (right) graphene and BN stripe bisects C-B (N-C) bonds. Here the difference charge density corresponding to a given applied field $\vec{E_y}$ is calculated as

\begin{equation}
\Delta \rho(\textbf{r},\vec{E_y})= \rho(\textbf{r},\vec{E_y})- \rho(\textbf{r},\vec{E_y}=0)
\end{equation}
where the total charge density of the PNDC under $\vec{E_y}$=0 is subtracted from the total charge density under the applied $\vec{E_y}$, and

\begin{equation}
\rho{(\textbf{r}, \vec{E_y})}= -e \sum^{occ}_{n,k,\sigma} \Psi^{*}_{n,k,\sigma}(\textbf{r},\vec{E_y})\Psi_{n,k,\sigma}(\textbf{r},\vec{E_y}).
\end{equation}

Here, $\Psi_{n,k,\sigma} $ indicates the spin polarized electronic state.
The same analysis was also performed for the right graphene stripe which is charged positively. It should be noted that the total charge density, $\rho(\textbf{r},\vec{E_y})$ integrated over the entire unit cell has to be zero, which has been verified also for each value of applied $\vec{E_y}$.

According to the atomic model presented in Fig. \ref{fig1} and Fig. \ref{fig2}, the states of two graphene stripes have small coupling and hence their bands are split in momentum space. The evolution of the energy bands with applied electric field $\vec{E_y}$ is presented for 0.2 V/\AA $\leq E_y \leq$ 1.0V/\AA~ in Fig. \ref{fig2}(a). Two unoccupied spin down bands (indicated with small arrows) shift in opposite directions, where the one associated with the left graphene stripe moves down and starts to become occupied. Concomitantly, a partially occupied spin up band associated with the right graphene stripe rises and hence becomes gradually depopulated. This situation explains how the charge and spin separations are achieved. Interestingly, for $ E_y \sim  $ 0.6 V/\AA~ the PNDC becomes a \textit{half-metal} with $\mu$=2.0 $\mu_B$. Similar half-metallic behavior in charged graphene/BN composites were also reported before. \cite{hm1, hm2, hm3} The shorting of graphene stripes and hence the shorting of PNDC is hindered by the dielectric BN spacer between graphenes.

In Fig. \ref{fig2}(c) we present the difference charge density, $ \Delta \rho (\textbf{r},\vec{E_y})$ of the PNDC under the field of $E_y$=1 V/\AA. The electron transfer from the right graphene stripe to the left stripe and the charge separation is clearly seen. In Fig. \ref{fig2}(d), we show $(xz)$-plane averaged electronic potential $\bar{V}(y)$. The potential at the left graphene stripe is lowered relative to the right graphene stripe, which is in compliance with the normal operation of PNDC. The electronic potential difference between two graphene stripes is calculated as $\Delta \bar{V}$ = 3 V under $E_y$=1 V/\AA.

\begin{figure}
\includegraphics[width=8cm]{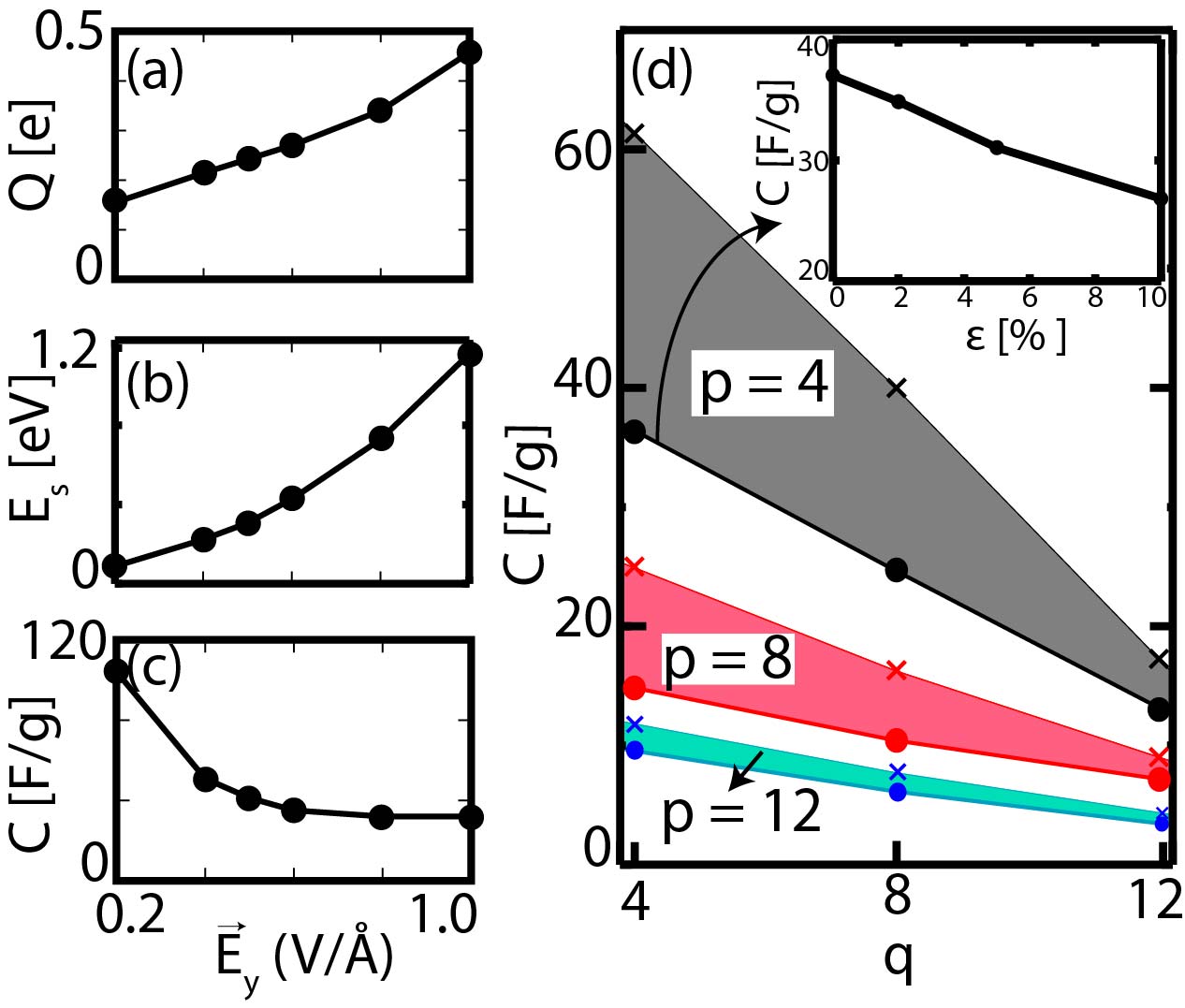}
\caption{ (a) Variation of excess charge $Q$ (e/cell); (b) stored energy $E_s$ (eV/cell) and (c) the corresponding gravimetric capacitance $C$ (F/g) for PNDC[4/4/4].  (d) Capacitance $C$ values in (F/g) of the planar nanoscale dielectric capacitor PNDC[p/q/p] are calculated for p=4-12 and q=4-12 for specific value of $E_y$ for which $C$ saturates. For each value of p, lower line connected by dots corresponds to capacitance values calculated through the expression, $C=Q^2/2mE_s$, while the upper line connected by crosses is computed from $C'=Q/m \Delta \bar{V}$. The calculated variation of $C$ with tensile strain, $\epsilon$ is shown by inset. }

\label{fig3}
\end{figure}

In Fig. \ref{fig3}(a-b) we present the variation of calculated charge $Q$ and stored energy $E_s$ as a function of $\vec{E_y}$. The energy stored on PNDC as a function of $\vec{E_y}$, is obtained from the expression

\begin{equation}
E_s(\vec{E_y})=E_{T}(\vec{E_y})-E_{T}(\vec{E_y}=0),
\end{equation}
where $E_{T}(\vec{E_y})$ is the total energy calculated for a given applied field $\vec{E_y}$. Using the values of $Q$ and $E_s$ calculated for a given $\vec{E_y}$, the capacitance per unit mass $C$ can be obtained from

\begin{equation}
C=Q^{2}/2mE_s,
\end{equation}
where $m$ is the mass of PNDC per unit cell. The variation of the capacitance as a function of $\vec{E_y}$ is shown in Fig. \ref{fig3}(c), which is high at small $\vec{E_y}$ (or bias voltage $V$) and gradually saturates to a constant value.

To explore the effect of size of PNDC[p/q/p] we repeated our analysis for different widths of graphene and BN stripes, namely for different p and q. Moreover, because of uncertainties involved in the effective widths of metallic and dielectric stripes (hence in the boundaries of the integral given in equation \ref{integral}), which affect the calculated values of $Q$ and $C$, here we present two alternative approaches in calculating $C$. In the first approach, $C$ is calculated from the expression $C=Q^{2}/2mE_s$, where $Q$ and $E_s$ correspond to $\vec{E_y}$ leading to the saturated values of $C$ as shown in Fig. \ref{fig3}(c). In the second approach, we followed the definition of the capacitance as $C'=Q/m \Delta \bar{V}$. Results obtained from these two alternative calculations are presented in Fig. \ref{fig3}(d), where the capacitances of PNDCs are plotted for different q values. For both approaches, $C$ decreases as q increases. Also $C$ decreases as the width of graphene stripe p, increases. However, the difference between the values of capacitance calculated by using two alternative approaches, namely $C'-C$  increases for small p and q, but decreases and eventually diminish as p and q get larger.

The capacitance values of PNDCs are found to depend strongly on the external conditions. For example, as shown by inset in Fig. \ref{fig3}(d), $C[4/4/4]$ changes by $\sim$ 30\% under the tensile strain of $\epsilon$=10\%, which can be detected easily and may offer technological applications. Similarly, when deformed under normal force or pressure, a PNDC can be deformed, whereby geometric  parameters such as the effective width of BN, $w$ is modified. Also, the energy barrier, $\Phi$ for electrons tunneling from left graphene to right graphene between two graphene stripes can be modified with local forces. Consequently, the capacitance value of PNDC is modified by local force or pressure. Capacitance value can also vary with the ambient temperature, since charge separation and hence $Q$, can be modified by the Fermi-Dirac distribution at finite temperature. Consequently, PNDCs can function as pressure and temperature sensors, when embedded in a soft materials like peptides or artificial tissues. It is noted that graphene/BN/graphene junctions of PNDC with small $w$ can work also as tunneling diodes or resonant tunneling device, where the tunneling current $I_t$ can be monitored by mechanical deformation, bias voltage or by temperature. Accordingly, these junctions also attain critical functionalities. Notably, all these features discussed above can be realized in diverse geometries, in particular in tubular forms.

Next we consider a series combination of PNDC[p/q/p] which is composed of commensurate sequential stripes, graphene(p)/BN(q)/graphene(2p)/BN(q)/graphene(p) stacked laterally. Accordingly, the middle graphene stripe has the twice width of the graphene stripes at either side of series PNDC. Under the applied electric field the excess negative and positive charges were accumulated at the far left and right edges of the middle graphene(2p). The capacitance of this series combination was calculated as half of the capacitance of PNDC[p/q/p], namely $C_{eq}=$ C[p/q/p] / 2. This demonstrates that to attain the desired $Q$ or $E_s$ or $\Delta V$ various combinations PNDC can be achieved on a single honeycomb structure by arranging metallic interconnects between parallel graphene stripes (which are graphene nanoribbons by themselves).

Finally, the PNDC structure revealed in this paper suggests us the construction of composite materials composed of the lateral stacking of a sequence of lattice matched zigzag ZGNR(p) / ZBNR(q) / ZGNR(p$^\prime$)... or armchair AGNR(p) / ABNR(q) / AGNR(p$^\prime$).. nanoribbons embedded on a large single layer, 2D honeycomb structure. For p=4 and q=4 this structure can be viewed as line compound consisting of the adjacent narrow stripes of graphene and BN, or $\delta$-doping for $p>>4$ and q=4 or vise versa. In the composite materials with sequence $p>>4$ and $q>>4$ the graphene stripes behave as if multichannel 1D metals on a 2D plane, which are separated by wide BN stripes. Depending on the values of p and q, these composites can display a variety of electronic structures and thus herald the engineering of a large class of materials.

\section{Conclusions}
In conclusion, we propose a planar nanoscale dielectric capacitor model which allow diverse parallel, series and mixed combinations to achieve desired values of charge separation, energy storage and potential difference values, and can be fabricated in the same plane of micro or nanocircuits for crucial electronic applications. They are flexible and can also form in tubular geometry to serve as pressure, heat and strain sensors. Present results let us also infer interesting 2D composite materials constructed from commensurate and lateral stacking of graphene and BN nanoribbons with zigzag or armchair edges.

\end{document}